\pdfoutput=1

\documentclass[nofootinbib,aps,prd,preprint,12pt,superscriptaddress]{revtex4}

\usepackage{mathrsfs}
\usepackage{hyperref}
\usepackage{amsmath,amsfonts}
\usepackage{color}
\usepackage{cancel}
\usepackage{multirow}
\usepackage{graphicx}

\makeindex
\renewcommand{\bar}{\overline}
\renewcommand{\hat}{\widehat}

\newcommand{\smallminus}{{\rm\rule[2.4pt]{6pt}{0.65pt}}}
\newcommand{\smallplus}{\hspace{0.5pt}\text{{\small+}}\hspace{-0.5pt}}
\newcommand{\eq}[1]{\begin{equation}#1\end{equation}}

\definecolor{paper_blue}{rgb}{0.2,0.1,0.75}

\definecolor{paper_blue}{rgb}{0.3,0.2,0.75}
\definecolor{paper_red}{rgb}{0.65,0.1,0.15}
\definecolor{ggreen}{rgb}{0.05,0.35,0.125}

\definecolor{paper_grey}{gray}{0.375}

\definecolor{downstairs}{rgb}{0.09, 0.1328, 0.7888}
\definecolor{upstairs}{rgb}{0.7598,0.1259,0.259}
\newcommand{\softdown}[1]{_{\raisebox{-1.95pt}{{\footnotesize{\color{downstairs}#1}}}}}
\newcommand{\softup}[1]{^{\raisebox{0.05pt}{{\footnotesize{\color{upstairs}#1}}}}}
\newcommand{\soft}[2]{\softup{#1}\softdown{#2}\!}

\newcommand{\beq}{\begin{equation}}
\newcommand{\eeq}{\end{equation}}
\newcommand{\beqa}{\begin{eqnarray}}
\newcommand{\eeqa}{\end{eqnarray}}

\begin{document}
\title{{~\\[-2cm]{\Large The Grassmannian and the Twistor String:\\Connecting All Trees in $\mathcal{N}=4$ SYM}}}

\author{Jacob L. Bourjaily}
\author{Jaroslav Trnka}
\affiliation{School of Natural Sciences, Institute for Advanced Study, Princeton, NJ 08540, USA}
\affiliation{Department of Physics, Princeton University, Princeton, NJ 08544, USA}
\author{Anastasia Volovich}
\author{Congkao Wen}
\affiliation{Department of Physics, Brown University, Providence, RI 02912, USA\\[1cm]~}
\begin{abstract}
We present a new, explicit formula for all tree-level amplitudes in $\mathcal{N}=4$ super Yang-Mills. The formula is written as a certain contour integral of the connected prescription of Witten's twistor string, expressed in link variables.
A very simple deformation of the integrand gives directly the Grassmannian integrand proposed in \cite{ArkaniHamed:2009dn} together with the explicit contour of integration.
The integral is derived by iteratively adding particles to the Grassmannian integral, one particle at a time, and makes manifest both parity and soft limits. The formula is shown to be related to that of \cite{DG}, and generalizes the results of \cite{NVW,ABCTtwo} for NMHV and N$^{2}$MHV to all N$^{(k-2)}$MHV tree amplitudes in $\mathcal{N}=4$ super Yang-Mills.

\end{abstract}

\maketitle

\tableofcontents\newpage

\section{Introduction}
There is now a vast and growing body of evidence to support the duality 
conjectured by Arkani-Hamed, Cachazo, Cheung and Kaplan \cite{ArkaniHamed:2009dn}
between the leading singularities\footnote{Leading singularities are $L$-loop integrals in field-theory evaluated along $T^{4L}$-contours which put $4L$ internal propagators on-shell.} of planar N$^{(k-2)}$MHV scattering amplitudes in $\mathcal{N}=4$ super Yang-Mills and certain contour integrals denoted $\mathcal{L}_{n,k}$ over the Grassmannian manifold $G(k,n)$ of $k$-planes in $n$-dimensions \cite{ArkaniHamed:2009dn,DG,SV,Mason:2009qx,ArkaniHamed:2009vw,Bullimore:2009cb,Kaplan:2009mh,ABCTone,NVW,ABCTtwo,Drummond:2010qh,Drummond:2010uq,Korchemsky:2010ut,DGtwo,Broedel:2010rr}. Parameterizing $G(k,n)$ in terms of a $k\times n$ matrix $C_{\alpha\,a}$---composed of $k$ representative vectors in $\mathbb{C}^n$ which span a given plane---$\mathcal{L}_{n,k}$ is given by
\eq{\label{Lnk1}\mathcal{L}_{n,k}=\frac{1}{\mathrm{vol[GL}(k)]}\oint\limits_{\Gamma_{n,k}}\frac{d^{k\times n}C_{\alpha\,a}}{(1)(2)(3)\cdots(n-1)(n)}\prod_{\alpha=1}^k\delta^{4|4}\left(C_{\alpha\,a}\mathcal{W}_{a}\right),} where $a=1,\ldots,n$ labels each particle, each $\mathcal{W}_a\equiv(\tilde{\mu},\tilde{\lambda}|\tilde{\eta})_a$ denotes a supertwistor which encodes the external momenta and helicities, and `$(j)$' represents the $j^{\mathrm{th}}$ $k\times k$-minor of $C_{\alpha\,a}$ built out of {\it consecutive} columns of the matrix $C_{\alpha\,a}$,\footnote{We will often use a single number---e.g.  `$(1)$'---to denote a {\it consecutive} minor beginning with the indicated column. More generally, a $k\times k$ minor constructed out of columns $\left[\ell_1,\ldots,\ell_k\right]$ $C_{\alpha\,a}$ will be denoted $(\ell_1\ldots \ell_k)$.} \eq{(j)\equiv(j\,\,j\smallplus1\,\,\cdots\,\,j\smallplus k\smallminus1)\equiv\epsilon^{\alpha_1\,\alpha_2\,\cdots\,\alpha_k}C_{\alpha_1\,j}C_{\alpha_2\,j+1}\cdots C_{\alpha_k\,j+k-1}.} 

Of course, as a contour integral, equation (\ref{Lnk1}) is nothing but the sum of the residues of the poles `encompassed' by the contour of integration $\Gamma_{n,k}$. The combinations of residues which compute tree amplitudes can be obtained by a variety of field-theoretic techniques, including the BCFW recursion relations \cite{Britto:2004ap, Britto:2005fq} (which can be efficiently translated in terms of the residues of $\mathcal{L}_{n,k}$, \cite{Kaplan:2009mh,Bullimore:2009cb,ABCCKT:2010}). It was not until recently, however, that the contours $\Gamma_{n,k}$ which compute tree amplitudes in $\mathcal{L}_{n,k}$ were understood in a way purely intrinsic to the Grassmannian. This understanding made manifest a deep connection between the Grassmannian integral $\mathcal{L}_{n,k}$ and Witten's twistor string theory. Because this connection is crucial to our main result, we briefly review it here before presenting our proposal for the contours which give all tree amplitudes in $\mathcal{N}=4$ super Yang-Mills.

Amplitudes in Witten's twistor string theory \cite{Witten:2003nn} can be computed via the `connected
prescription' written down by Roiban, Spradlin and one of the authors in \cite{Roiban:2004yf,RSV,Roiban:2004ka} as integrals of an open string correlator over
the moduli space of curves in supertwistor space. Although geometrically very beautiful, these integrals turned out to be technically very difficult to use because of the presence of highly non-linear equations. Using the link variables described in \cite{SV,DG}, Dolan and Goddard \cite{DG} wrote contour integrals which compute all tree amplitudes as rational functions, and checked explicitly that these lead to the correct formulae for many particular amplitudes
(see also \cite{SV}), and for all split-helicity amplitudes in \cite{DGtwo}. The key insight of Dolan and Goddard was to use a sequence of global residue theorems\footnote{The global residue theorem is the multi-dimensional generalization of Cauchy's theorem of ordinary contour integrals in one complex dimension, called the global residue theorem (see, e.g. \cite{Griffiths:1978a}).} which connect the connected prescription contours to $\mathcal{L}_{n,k}$. Significantly, the twistor string construction---especially when expressed in the framework of the connected prescription---carries with it the knowledge of a natural, preferred choice of integration contour which computes each tree amplitude. The particle interpretation in $\mathcal{L}_{n,k}$ made this tractable. 

The equivalence between the connected prescription for the twistor string and $\mathcal{L}_{n,k}$ was recently proven for all NMHV amplitudes in \cite{NVW,ABCTtwo}. These proofs rely on repeated use of the global residue theorem, and show that the combination of residues contributing to any NMHV amplitude
computed via the twistor string can be re-expressed as a direct sum
of residues of $\mathcal{L}_{n,k}$.
Moreover, an amazing and much stronger property was observed:
the two integrands were in fact related by a {\it smooth deformation}, which interpolates between the connected prescription of twistor string theory and the Grassmannian integrand $\mathcal{L}_{n,k}$. The deformation connecting the two descriptions moves the locations of each pole, and changes the value of each residue;
but the sum of residues which define the tree amplitude is itself found to be invariant.   
Taking together the results of \cite{NVW, ABCTtwo}, that the
twistor string connected prescription provides a preferred
choice of integration contour and that its integrand may be
smoothly deformed to the integrand $\mathcal{L}_{n,k}$, we conclude
therefore that the twistor string may be used to generally answer
the important open question of determining the appropriate contours in the Grassmannian for computing any general tree amplitude in $\mathcal{N}=4$ super Yang-Mills. For this to be the case, it is necessary that the contour given for the connected prescription continue to make manifest the connection between the twistor string and the Grassmannian through a contour deformation similar to that described in \cite{NVW, ABCTtwo} for NMHV amplitudes. 

In this paper, we propose a new, explicit formula for all N$^{(k-2)}$MHV tree amplitudes in $\mathcal{N}=4$,
generalizing the NMHV results of \cite{NVW, ABCTtwo}.
In section 2 we will present our main formula, \mbox{equation (\ref{general_formula})}, and discuss its smooth deformation to a contour in $\mathcal{L}_{n,k}$.
In section 3 we will describe how this formula can be obtained by iteratively `adding particles' in a natural way to the first non-trivial tree amplitude, the $6$-point NMHV amplitude, while making sure that soft limits and parity are manifest at every stage.
In section 4 we will make a series of transformations to map our formula to that of \cite{DG},
thus deriving it from twistor string connected prescription.

\newcommand{\flist}{\sigma}

\section{All Tree Amplitudes in $\mathcal{N}=4$ Super Yang-Mills}
We propose that the general, tree-level, planar, color-stripped, $n$-point N$^{(k-2)}$MHV amplitude can be written
\eq{\label{general_formula}
\hspace{-0.cm}\mathscr{A}_{n}^{(k)}=\frac{1}{\mathrm{vol[GL(}k\mathrm)]}\oint\limits_{\mathscr{F}_n^{(k)}=\vec{0}} \hspace{-0.15cm}\frac{dC_{\alpha\,a}\,\,\,\mathscr{H}_{n}^{(k)}}{(n-1)(1)(3)
\,\,\mathscr{F}_{n}^{(k)}}\,\,\prod_{\alpha=1}^k\delta^{4|4}\left(C_{\alpha\,a}\mathcal{W}_{a}\right),
}
where the contour $\mathscr{F}_n^{(k)}=\vec{0}$ is the zero-locus of $\mathscr{F}_n^{(k)}:\mathbb{C}^{(n-k-2)(k-2)}\to\mathbb{C}^{(n-k-2)(k-2)}$, defined in terms of the \mbox{$(n-k-2)(k-2)$} Veronese maps $F_\ell^j$,
\eq{\label{def_of_F}\mathscr{F}_n^{(k)}\equiv\prod_{\ell=k+3}^n\left(\prod_{j=1}^{k-2}F_{\ell}^j\right),} where each $F_{\ell}^j$ can be written in terms of the minors of $C_{\alpha\,a}$ according to \eq{\begin{split}F_\ell^j\equiv&\phantom{\,-\,}\left(\flist_{\ell}^j\,\,\,\,\ell\smallminus2\,\,\ell\smallminus1\,\,\ell\right)\left(\flist_{\ell}^j\,\,\,\,\ell\,\,j\,\,j\smallplus1\right)\left(\flist_{\ell}^j\,\,\,\,j\smallplus1\,\,j\smallplus2\,\,\ell\smallminus2\right)\left(\flist_{\ell}^j\,\,\,\,\ell\smallminus1\,\,j\,\,j\smallplus2\right)\\&-\left(\flist_{\ell}^j\,\,\,\,j\,\,j\smallplus1\,\,j\smallplus2\right)\left(\flist_{\ell}^j\,\,\,\,j\smallplus2\,\,\ell\smallminus2\,\,\ell\smallminus1\right)\left(\flist_{\ell}^j\,\,\,\,\ell\smallminus1\,\,\ell\,\,j\right)\left(\flist_{\ell}^j\,\,\,\, j\smallplus1\,\,\ell\smallminus2\,\,\ell\right),\end{split}\label{explicit_veronese_maps}}
with $\flist_{\ell}^j$ representing collectively the columns $\left[1,\ldots, j\smallminus1\right]\cup\left[j\smallplus\ell\smallminus k,\ldots,\ell\smallminus3\right]$ of $C_{\alpha\,a}$, and where $\mathscr{H}_n^{(k)}$ is the product of all the {\it non-consecutive} minors in the {\it first line} of \mbox{equation (\ref{explicit_veronese_maps})}; \nopagebreak explicitly,
\begin{align}\vspace{-0.6cm}\hspace{-0.0cm}\mathscr{H}_n^{(k)}=&\,\,\mathscr{H}_{n-1}^{(k)}\times(\flist_{n-1}^{k-2}\,\,n\smallminus1\,\,k\smallminus2\,\,k\smallminus1)\nonumber\\[-0.1cm]&\times\prod_{j=1}^{k-3}\left[(\flist_n^j\,\,n\,\,j\,\,j\smallplus1)(\flist_{n-1}^{j+1}\,\,n\smallminus3\,\,n\smallminus2\,\,n\smallminus1)\right]\prod_{j=1}^{k-2}\left[(\flist_{n}^j\,\,n\smallminus1\,\,j\,\,j\smallplus2)(\flist_n^j\,\,j\smallplus1\,\,j\smallplus2\,\,n\smallminus2)\right].\nonumber\vspace{-0.6cm}\end{align} 
Noticing that all the minors appearing in a given map $F_{\ell}^j$ involve the same set of columns $\sigma_{\ell}^j$, and that the rest are organized according to a `$3\times 3$' Veronese operator, we may encode the structure of \mbox{equation (\ref{explicit_veronese_maps})} by writing\footnote{This simplified notation can be justified by observing that only $6$ of the $k+3$ columns which are relevant to a given Veronese operator $F_\ell^j$ change from one term to another.} 
\eq{\vspace{-0.3cm} \label{F} \begin{split}F_\ell^j\equiv& \flist_{\ell}^{j}\bowtie S_{\ell-2\,\,\ell-1\,\,\ell\,\,j\,\,j+1\,\,j+2},\\\equiv&\,\,\left[1,\ldots, j\smallminus1;\,\,\,\,j\smallplus\ell\smallminus k,\ldots,\ell\smallminus3\right]\bowtie S_{\ell-2\,\,\ell-1\,\,\ell\,\,j\,\,j+1\,\,j+2},\end{split}}
where $S_{a\,b\,c\,d\,e\,f}$ represents the primitive Veronese operator which, when acting on $\mathbb{P}^2$, tests if the six points $a,\ldots, e$ lie on a conic,
\eq{S_{a\,b\,c\,d\,e\,f}\equiv\,\,(a\,b\,c)(c\,d\,e)(e\,f\,a)(b\,d\,f)-(b\,c\,d)(d\,e\,f)(f\,a\,b)(c\,e\,a).}

As will be described below, the structure of the numerators $\mathscr{H}_n^{(k)}$ is dictated by the proposed duality between \mbox{equation (\ref{general_formula})} and a related expression  in $\mathcal{L}_{n,k}$. Following the theme of \cite{ABCTtwo}, 
let us introduce a deformation parameter $t_{\ell}^j$ for each map $F_{\ell}^j$,
\eq{\label{deformed_veronese}\begin{split}F_\ell^j(t_\ell^j)\equiv&\phantom{\,-\,t_\ell^j\,}\left(\flist_{\ell}^{j}\,\,\,\,\ell\smallminus2\,\,\ell\smallminus1\,\,\ell\right)\left(\flist_{\ell}^{j}\,\,\,\,\ell\,\,j\,\,j\smallplus1\right)\left(\flist_{\ell}^{j}\,\,\,\,j\smallplus1\,\,j\smallplus2\,\,\ell\smallminus2\right)\left(\flist_{\ell}^{j}\,\,\,\,\ell\smallminus1\,\,j\,\,j\smallplus2\right)\\&-t_\ell^j\left(\flist_{\ell}^{j}\,\,\,\,j\,\,j\smallplus1\,\,j\smallplus2\right)\left(\flist_{\ell}^{j}\,\,\,\,j\smallplus2\,\,\ell\smallminus2\,\,\ell\smallminus1\right)\left(\flist_{\ell}^{j}\,\,\,\,\ell\smallminus1\,\,\ell\,\,j\right)\left(\flist_{\ell}^{j}\,\,\,\, j\smallplus1\,\,\ell\smallminus2\,\,\ell\right).\end{split}}
Then the integral $\mathscr{A}_n^{(k)}(t_\ell^j)$, with all $F_\ell^j$ in (\ref{general_formula})  replaced by $F_\ell^j(t_\ell^j)$, will map precisely to the one appearing for $\mathcal{L}_{n,k}$ in limit of $t_\ell^j\to0$ for all $\ell,j$. This is because, together with the three minors manifest in \mbox{equation (\ref{general_formula})} (namely, $(n-1)$, $(1)$, and $(3)$) the factors which form $\mathscr{F}_n^{(k)}(t_\ell^j)$ when $t_\ell^j \to 0$ will contribute exactly one copy of each of the consecutive minors present in the measure of the integral $\mathcal{L}_{n,k}$: 
\vspace{-0.25cm}\eq{\mathscr{F}_n^{(k)}=\underbrace{\Big(F_{k+3}^1\cdots F_{k+3}^{k-2}\Big)}_{\substack{\cup\\(2),(4)}}\underbrace{\Big(F_{k+4}^1\cdots F_{k+4}^{k-2}\Big)}_{\substack{\cup\\(5)}}\underbrace{\Big(F_{k+5}^1\cdots F_{k+5}^{k-2}\Big)}_{\substack{\cup\\(6)}}\cdots\underbrace{\Big(F_{n-1}^1\cdots F_{n-1}^{k-2}\Big)}_{\substack{\cup\\(n-k)}}\underbrace{\Big(F_{n}^1\cdots F_{n}^{k-2}\Big)}_{\substack{\cup\\(n-k+1),\ldots,(n-2),(n)}}.\nonumber\vspace{-0.2cm}}
And since $\mathscr{H}^{(k)}_n$ is composed of all the {\it non-consecutive} minors present in the {\it first} factors of each $F_\ell^j$, we have that 
\eq{\label{Lnklimit}\lim_{t_\ell^j\to0}\left(\frac{\mathscr{H}_n^{(k)}}{(n-1)(1)(3)\,\,\,\mathscr{F}_n^{(k)}}\right)=\frac{1}{(n-1)(1)(3)}\frac{1}{(2)\,\,(4)(5)\cdots(n-3)(n-2)\,(n)},} making the connection between the twistor string and $\mathcal{L}_{n,k}$ manifest. 

We strongly suspect that formula (\ref{general_formula}) is unchanged by any of the deformations introduced by the parameters $t_\ell^j$ in (\ref{deformed_veronese}). For NMHV amplitudes, $t_\ell^j$-independence has been rigorously proven by a direct application of the global residue theorem, \cite{ABCTtwo,NVW}, and we suspect that similar arguments can be used to prove $t_\ell^j$-independence more generally. We have checked this numerically for several nontrivial N$^2$MHV amplitudes, including for the alternating-helicity amplitude for eight gluons, but will leave the question of proving complete $t_\ell^j$-independence to future researches. 

Let us end this section by presenting explicitly the $t_\ell^j\to0$ limit of the deformed twistor-string contour (\ref{general_formula}), illustrating some of the key differences between the two formulations. When $t_\ell^j\to0$, each Veronese operator factorizes into the product of the four minors listed in the first line of (\ref{deformed_veronese}). In general, all but $n-3$ of these factors will be non-consecutive, and therefore are included among the factors of the numerator $\mathscr{H}_n^{(k)}$. Although it is generally ill-advised to `cancel terms' between the contour-defining maps defining $\mathscr{F}_n^{(k)}$ and the numerator, there is a good physical reason for suspecting that the `fourth' minors of each of the $F_\ell^j(t_\ell^j\to0)$---which are never consecutive---contribute no non-vanishing residues to the contour.\footnote{The reason why na\"ive cancellation of factors between $\mathscr{H}_{n}^{(k)}$ and those in $\mathscr{F}_{n}^{(k)}(t_\ell^j\to0)$ can be misleading is described with several examples in \cite{ABCTtwo}; for example, even the poles supported by purely non-consecutive minors of the $F_\ell^j$'s can have the interpretation of being supported by consecutive minors, and thereby contributing a residue to the contour.} As described in \cite{ABCTone,ABCTtwo}, CSW operators, when translated into the Grassmannian, are all constructed from products of three minors. Although beyond the scope of the present discussion, ensuring that each pole of the integrand is composed of three-minor operators helps one to connect the CSW, or `disconnected', support of tree amplitudes to the `connected' support of the twistor string through a series of global residue theorems. At any rate, there is now enough direct evidence that general tree-contours are entirely supported on the vanishing first three factors of each $F_\ell^j$ when $t_\ell^j\to0$ to justify the simplification to a `$3$-minor' form of each map in the contour.

Taking each $t_\ell^j \to 0$, the twistor-string contour $\mathscr{A}_{n}^{(k)}(t_l^j)$ becomes,\eq{\label{Lnk_general_formula}\mathscr{A}_n^{(k)}(t_\ell^j)\xrightarrow[t_\ell^j\to0]{}{}\mathcal{A}_n^{(k)}=\frac{1}{\mathrm{vol[GL(}k\mathrm{)]}}\oint\limits_{\mathcal{F}_n^{(k)}=\vec{0}}\frac{dC_{\alpha\,a}\,\,\,\mathcal{H}_{n}^{(k)}}{(n-1)(1)(3)
\,\,\mathcal{F}_{n}^{(k)}}\,\,\prod_{\alpha=1}^k\delta^{4|4}\left(C_{\alpha\,a}\mathcal{W}_{a}\right),}
where 
\eq{\label{Lnk_general_formula2}
\mathcal{F}_n^{(k)}\equiv\prod_{\ell=k+3}^n\left(\prod_{j=1}^{k-2}f_{\ell}^j\right)~~{\rm with}~~
f_\ell^j\equiv \sigma_{\ell}^j  \bowtie \left(\ell\smallminus2\,\,\ell\smallminus1\,\,\ell\right)\left(\ell\,\,j\,\,j\smallplus1\right)\left(j\smallplus1\,\,j\smallplus2\,\,\ell\smallminus2\right),} with $\flist_\ell^j$ as before, and for where
\eq{
\mathcal{H}_{n}^{(k)}=
\frac{\mathscr{H}_n^{(k)}}
{\prod_{\ell=k+3}^n  \prod_{j=1}^{k-2} \left(\sigma_\ell^j\,\,\ell\smallminus1\,\,j\,\,j\smallplus2 \right) },}
which, as before, represents the product of all non-consecutive minors among the maps $f_\ell^j.$ 

Alternatively, we could have started with formula (\ref{Lnk_general_formula}) 
for $\mathcal{A}_{n}^{(k)}$ and obtained formula (\ref{general_formula}) for
$\mathscr{A}_{n}^{(k)}$
by ``adding a missing minor" to each 
\eq{\label{deformed_veronese2} \begin{split} f&=\sigma \bowtie(a\,b\,c)(c\,d\,e)(e\,f\,a)\\ \Rightarrow F&=\sigma \bowtie\left[ (a\,b\,c)(c\,d\,e)(e\,f\,a)(b\,d\,f)-(b\,c\,d)(d\,e\,f)(f\,a\,b)(c\,e\,a) \right], \end{split}}
in order to supply a simple geometric meaning to the contour---the maps $F$'s having the natural interpretation of testing the localization of points in $\mathbb{P}^{(k-1)}$.

Both formulae give all tree-level amplitudes in $\mathcal{N}=4$ super Yang-Mills in terms of a particular contour integral. The first one, (\ref{general_formula}), naturally arises from the twistor string theory. Its contour $\mathscr{F}_n^{(k)}=\vec{0}$ has a nice geometric meaning: it is the constraint for $n$ points to lie on a degree-$(k-1)$ curve in twistor space. On the other hand, the formula (\ref{Lnk_general_formula}) 
provides the integration contour for Grassmannian ${\cal L}_{n,k}$, and thereby ensures that each contribution is itself manifestly Yangian invariant.

\newcommand{\AnkInt}[4]{\frac{1}{\mathrm{vol[GL(}#2\mathrm{)]}}\oint\limits_{\mathscr{F}_{#1}^{(#2)}=\vec{0}}\!\!\!\!\!dC_{\alpha\,a}\frac{#3}{#4}\prod_{\alpha=1}^{#2}\delta^{4|4}(C_{\alpha\,a}\mathcal{W}_a)}

\section{Building the General Contour, one Particle at a Time}
In this section we describe how the general contour for any tree amplitude (\ref{general_formula}) can be obtained by sequentially extending the contour of the first non-trivial amplitude, the $6$-point NMHV amplitude, by adding one particle at a time. Before doing so, however, it will be useful to briefly discuss some of the generally-desirable features that any such contour-prescription should have.

Let us consider what would be necessary to extend a formula valid for $\mathcal{L}_{n-1,k}$ to one valid for $\mathcal{L}_{n,k}$ for some fixed $k$. Recall that the integral $\mathcal{L}_{n,k}$'s measure is given by the product of the $n$ consecutive $k\times k$ minors of $C_{\alpha\,a}$. The $n^{\mathrm{th}}$ particle, being represented by the $n^{\mathrm{th}}$ column of $C_{\alpha\,a}$ participates in $k$ of these consecutive minors; and these $k$ minors, taken together, span a range of $\min(n,2k-1)$ columns of $C_{\alpha\,a}$. This suggests that, fixing $k$, only for $n\geq 2k-1$ will a tree contour be sufficiently general to have an extension to all $n$. Conveniently, the \mbox{$n=(2k-1)$-point} N$^{k-2}$MHV amplitude, $\mathscr{A}_{n=2k-1}^{(k)}$, is nothing but the parity-conjugate of the \mbox{$n$-point} N$^{k-3}$MHV amplitude, $\mathscr{A}_{n=2k-1}^{(k-1)}$, allowing it to be uniquely related to a contour with strictly lower-$k$. And so it should be possible to `bootstrap' a formula valid for any fixed $k$ to one valid for all $k$, using parity when $n=2k-1$ as the bridge which connects each \mbox{$k$ to $k+1$}.  

Just as there are several equally-valid formulae for the general NMHV tree contour \mbox{(see, e.g. \cite{ABCTtwo,DG,NVW,SV})}, there are several ways of writing the general N$^{(k-2)}$MHV tree contour. The one that we derive here is obtained by starting with the particular NMHV tree contour given in \cite{ABCTtwo} and extending it in such a way that the general contour prescription is invariant under parity for all $n,k$. As we will see, these criteria lead uniquely to the contour given here which defines our general result (\ref{general_formula}).\footnote{We have also found other parity-symmetric contour prescriptions by starting from each of the  different forms of the NMHV tree amplitude. We have checked that each of these extensions to all $n,k$ is unique and that each leads to correct formulae for general tree amplitudes. In addition, there are further possibilities if one foregoes the connection between $\mathcal{L}_{n,k}$ and the twistor string, but those will not be considered here.} 

\subsection{NMHV amplitudes}
Let us begin with the simplest amplitude which requires a non-trivial contour to be specified. The $6$-point tree amplitude's contour is essentially unique up to a global residue theorem, and can be written \cite{ABCTtwo,DG,ArkaniHamed:2009dn,NVW,SV},
\eq{\mathscr{A}_{6}^{(3)}=\AnkInt{6}{3}{\mathscr{H}_6^{(3)}}{(5)(1)(3)\,\,\,\mathscr{F}_{6}^{(3)}},\vspace{-0.4cm}}
where\vspace{-0.3cm}\eq{\begin{split}\mathscr{F}_6^{(3)}=&\Big[{\color{paper_red}(4)(6)(2)}{\color{paper_grey}(1\,3\,5)-(5\,6\,1)(1\,2\,3)(3\,4\,5)(6\,2\,4)}\Big]=S_{4\,5\,6\,1\,2\,3}\\\mathrm{and}\qquad\mathscr{H}_{6}^{(3)}=&\,\,\qquad\qquad{\color{paper_grey}(1\,3\,5)}.\end{split}}
(Here, we have chosen to de-emphasize the minors which do not appear in the analogous expressions for $\mathcal{L}_{n,k}$ by colouring them grey. We have also chosen to highlight all the {\it consecutive} minors which participate in the contour by colouring them red; this highlighting will be useful when we consider amplitudes involving more particles and with \mbox{$k>3$}.)

As demonstrated in \cite{ABCTtwo}, this contour can be extended to all NMHV amplitudes in the following way, 
\eq{\label{general_nmhv}\mathscr{A}_n^{(3)}=\AnkInt{n}{3}{\prod_{\ell=6}^{n-1}\big[(1\,2\,\ell)(2\,3\,\ell\smallminus1)\big]\prod_{\ell=6}^n\big[{\color{paper_grey}(1\,3\,\ell\smallminus1)}\big]}{(n-1)(1)(3)\,\,\,\mathscr{F}_{n}^{(3)}},\vspace{-0.4cm}}
where \eq{\hspace{-1.5cm}\mathscr{F}_n^{(k)}=\prod_{\ell=6}^n\Big[{\color{paper_red}(\ell\smallminus2\,\,\ell\smallminus1\,\,\ell)}(\ell\,\,1\,\,2)(2\,\,3\,\,\ell\smallminus2){\color{paper_grey}(\ell\smallminus1\,\,1\,\,3)\,-\,(\ell\smallminus1\,\,\ell\,\,1)(1\,\,2\,\,3)(3\,\,\ell\smallminus2\,\,\ell\smallminus1)(\ell\,\,2\,\,\ell\smallminus1)}\Big]=\prod_{\ell=6}^nS_{\ell-2\,\,\ell-1\,\,\ell\,\,1\,\,2\,\,3}.\nonumber}

Notice that the only operator that involves particle $n$ is the last, $F_{\ell=n}^{j=1}$, and this operator includes in general all but one of the consecutive minors which involve $n$---namely, all but minor $(n-1)$. Indeed, each $F_{\ell}^1$ can be seen as an operator which adds particle $\ell$ to the $(\ell-1)$-point contour. 

Consider for example the contour for $n=7$,
\eq{\hspace{-0.05cm}\mathscr{F}_7^{(3)}=\begin{tabular}{rlrcccll}\multirow{1}{*}{$\left\{\rule[-2pt]{0pt}{12.5pt}\right.$}&$F_6^1=\!\!\!$&$\!\!\;\;\;{\color{paper_red}(4)}\;\;\;\;{\color{black}(6\,1\,2)}\;\;\;\;{\color{paper_red}(2)}\;\;\;{\color{paper_grey}(5\,1\,3)}$&$-$&${\color{paper_grey}(5\,6\,1)}{\color{paper_grey}(1\,2\,3)}{\color{paper_grey}(3\,4\,5)}{\color{paper_grey}(6\,2\,4)}$&$=S_{4\,5\,6\,1\,2\,3}$&\multirow{1}{*}{$\left.\rule[-2pt]{0pt}{12.5pt}\right\}$}\\\multirow{1}{*}{$\left\{\rule[-2pt]{0pt}{12.5pt}\right.$}&$F_7^1=\!\!\!$&$\!\!\;\;{\color{paper_red}(5)}\;\;\;\;\;\;\;{\color{paper_red}(7)}\;\;\;\;{\color{black}(2\,3\,5)}{\color{paper_grey}(6\,1\,3)}$&$-$&${\color{paper_grey}(6\,7\,1)}{\color{paper_grey}(1\,2\,3)}{\color{paper_grey}(3\,5\,6)}{\color{paper_grey}(7\,2\,5)}$&$=S_{5\,6\,7\,1\,2\,3}$&\multirow{1}{*}{$\left.\rule[-2pt]{0pt}{12.5pt}\right\}$}\end{tabular}.}
By recognizing that $\mathscr{A}_7^{(3)}$ is nothing but the parity-conjugate of $\mathscr{A}_{7}^{(4)}$, we may use this contour to directly obtain the contour of the first non-trivial N$^2$MHV tree-amplitude. 

\subsection{N$^2$MHV Amplitudes}

As mentioned above, because the parity-conjugate\footnote{Here, we should point out that we are using a definition of `parity' that both exchanges the column-labels of each minor with the complement of each, and maps each column \mbox{$j\mapsto(n+1)-j$}. This appears to be the most natural definition of parity in the Grassmannian.} of the $7$-point NMHV amplitude is the $7$-point N$^2$MHV amplitude, we may use the general NMHV contour to obtain our first non-trivial contour for $k=4$,
\eq{\hspace{-1.5cm}\mathscr{F}_7^{(4)}=\widetilde{\mathscr{F}_7^{(3)}}=\begin{tabular}{rlrcccll}\multirow{2}{*}{$\left\{\rule[-17pt]{0pt}{35pt}\right.$}&$F_7^1=\!\!\!$&$\!\!\;\;\;{\color{paper_red}(4)}\;\;{\color{black}(4\,7\,1\,2)}\;\;\;\;{\color{paper_red}(2)}\;\;\;{\color{paper_grey}(4\,6\,1\,3)}$&$-$&${\color{paper_grey}(4\,1\,2\,3)}{\color{paper_grey}(4\,3\,5\,6)}{\color{paper_grey}(4\,6\,7\,1)}{\color{paper_grey}(4\,2\,6\,8)}$&$=\left[4\right]\bowtie S_{5\,6\,7\,1\,2\,3}$&\multirow{2}{*}{$\left.\rule[-17pt]{0pt}{35pt}\right\}$}\\&$F_7^2=\!\!\!$&$\!\!\;\;\;{\color{paper_red}(5)}\;\;\;\;\;{\color{paper_red}(7)}\;\;\;\;{\color{black}(1\,{\color{paper_blue}3\,4\,5})}{\color{paper_grey}(1\,6\,2\,4)}$&$-$&${\color{paper_grey}(1\,2\,3\,4)}{\color{paper_grey}(1\,4\,5\,6)}{\color{paper_grey}(1\,6\,7\,2)}{\color{paper_grey}(1\,3\,5\,7)}$&$=\left[1\right]\bowtie S_{5\,6\,7\,2\,3\,4}$\end{tabular}.\nonumber}

From here, there are several ways in which the above contour can be extended to one for all $n$. For example, one could make the identification made in \cite{ABCTtwo}, that \eq{\label{n2mhv_contour_1}F_7^{1,2}=\left\{\begin{array}{l}\left[4\right]\bowtie S_{5\,6\,7\,1\,2\,3}\\\left[1\right]\bowtie S_{5\,6\,7\,2\,3\,4}\end{array}\right\}\quad\Longrightarrow\quad F_{\ell}^{1,2}\Leftrightarrow\left\{\begin{array}{r}\left[\ell\smallminus3\right]\bowtie S_{\ell-2\,\,\ell-1\,\,\ell\,\,1\,\,2\,\,3}\\\left[1\right]\bowtie S_{\ell-2\,\,\ell-1\,\,\ell\,\,2\,\,3\,\,\ell-3}\end{array}\right\}.} However, this extension of the $7$-point N$^2$MHV amplitude leads to a form of the $8$-point N$^2$MHV contour which is not manifestly self-conjugate under parity, and which unnecessarily obfuscates the extension to all N$^{(k-2)}$MHV amplitudes.\footnote{That being said, we have every reason to suspect the formula given in \cite{ABCTtwo} is in fact equivalent to the one we present here.} We suggest that the following extension is more natural, 
\eq{\label{n2mhv_contour_2}F_7^{1,2}=\left\{\begin{array}{l}\left[4\right]\bowtie S_{5\,6\,7\,1\,2\,3}\\\left[1\right]\bowtie S_{5\,6\,7\,2\,3\,4}\end{array}\right\}\quad\Longrightarrow\quad F_{\ell}^{1,2}\Leftrightarrow\left\{\begin{array}{r}\left[\ell\smallminus3\right]\bowtie S_{\ell-2\,\,\ell-1\,\,\ell\,\,1\,\,2\,\,3}\\\left[1\right]\bowtie S_{\ell-2\,\,\ell-1\,\,\ell\,\,2\,\,3\,\,4}\end{array}\right\}.} Notice that the only difference between the contour prescriptions in (\ref{n2mhv_contour_1}) and (\ref{n2mhv_contour_2}) is that the former associates $S_{5\,6\,7\,2\,3\,4}$ with $S_{\ell-2\,\ell-1\,\ell\,2\,3\,\ell-3}$ while the latter associates $S_{5\,6\,7\,2\,3\,4}$ with $S_{\ell-2\,\ell-1\,\ell\,2\,3\,4}$.

Using this prescription, we find that the 8-point N$^2$MHV may be written,
\eq{
\hspace{-0.cm}\mathscr{A}_{8}^{(4)}=\frac{1}{\mathrm{vol[GL(}4\mathrm)]}\oint\limits_{\mathscr{F}_{8}^{(4)}=\vec{0}} \hspace{-0.15cm}\frac{dC_{\alpha\,a}\,\,\,\mathscr{H}_{8}^{(4)}}{(7)(1)(3)
\,\,\mathscr{F}_{8}^{(4)}}\,\,\prod_{\alpha=1}^4\delta^{4|4}\left(C_{\alpha\,a}\mathcal{W}_{a}\right),
}
where $\mathscr{F}_{8}^{(4)}=F_7^1 F_7^2 \cdot F_8^1 F_8^2$ with the $F^j_{\ell}$ given explicitly by\footnote{Here, we have highlighted each of the primary `consecutive subparts' of each of the minors in the contour by colouring them blue. These tend to be the most important minors when computing a tree amplitude as a series of `geometry problems' as described in \cite{ABCTtwo}.}
\eq{\label{g48_contour}\hspace{-1.25cm}\mathscr{F}_8^{(4)}=\small
\begin{tabular}{rrrcccll}\multirow{2}{*}{$\left\{\rule[-17pt]{0pt}{35pt}\right.$}&$F_7^1=$&$\;\;{\color{paper_red}(4)}\;\;\;\;\;{\color{black}(4\,7\,1\,2)}\;\;\;\;{\color{paper_red}(2)}\;\;\;{\color{paper_grey}(4\,6\,1\,3)}$&$-$&${\color{paper_grey}(4\,1\,2\,3)}{\color{paper_grey}(4\,3\,5\,6)}{\color{paper_grey}(4\,6\,7\,1)}{\color{paper_grey}(4\,2\,6\,8)}$&$=\left[4\right]\bowtie S_{5\,6\,7\,1\,2\,3}$&\multirow{2}{*}{$\left.\rule[-17pt]{0pt}{35pt}\right\}$}\\&$F_7^2=$&${\color{black}(1\,{\color{paper_blue}5\,6\,7})}{\color{black}({\color{paper_blue}1\,2\,3}\,7)}{\color{black}(1\,{\color{paper_blue}3\,4\,5})}{\color{paper_grey}(1\,6\,2\,4)}$&$-$&${\color{paper_grey}(1\,2\,3\,4)}{\color{paper_grey}(1\,4\,5\,6)}{\color{paper_grey}(1\,6\,7\,2)}{\color{paper_grey}(1\,3\,5\,7)}$&$=\left[1\right]\bowtie S_{5\,6\,7\,2\,3\,4}$\\ 
\multirow{2}{*}{$\left\{\rule[-17pt]{0pt}{35pt}\right.$}&$F_8^1=$&$\;\;{\color{paper_red}(5)}\;\;\;\;\;{\color{black}(5\,{\color{paper_blue}8\,1\,2})}{\color{black}(5\,6\,2\,3)}{\color{paper_grey}(5\,7\,1\,3)}$&$-$&${\color{paper_grey}(5\,1\,2\,3)}{\color{paper_grey}(5\,3\,6\,7)}{\color{paper_grey}(5\,7\,8\,1)}{\color{paper_grey}(5\,2\,6\,8)}$&$=\left[5\right]\bowtie S_{6\,7\,8\,1\,2\,3}$&\multirow{2}{*}{$\left.\rule[-17pt]{0pt}{35pt}\right\}$}\\&$F_8^2=$&${\;\; \color{paper_red}(6)}\;\;\;\;\;\;\;\;{\color{paper_red}(8)}\;\;\;\;{\color{black}(1\,3\,4\,6)}{\color{paper_grey}(1\,7\,2\,4)}$&$-$&${\color{paper_grey}(1\,2\,3\,4)}{\color{paper_grey}(1\,4\,6\,7)}{\color{paper_grey}(1\,7\,8\,2)}{\color{paper_grey}(1\,3\,6\,8)}$&$=\left[1\right]\bowtie S_{6\,7\,8\,2\,3\,4}$&\end{tabular}}
and $\mathscr{{H}}_{8}^{(4)}$ is the product of all {\it non-consective} minors of the first factors of the $F_\ell^j$'s,
\eq{\begin{split}\mathscr{{H}}_{8}^{(4)}=&\phantom{\,\times\,}{\color{black}(4\,7\,1\,2)}{\color{black}(1\,5\,6\,7)}{\color{black}(1\,2\,3\,7)}{\color{black}(1\,3\,4\,5)}{\color{black}(5\,8\,1\,2)}{\color{black}(5\,6\,2\,3)}{\color{black}(1\,3\,4\,6)}\\& \times {\color{paper_grey}(4\,6\,1\,3)}{\color{paper_grey}(1\,6\,2\,4)}{\color{paper_grey}(5\,7\,1\,3)}{\color{paper_grey}(1\,7\,2\,4)}\, .\end{split}}
It is not hard to see that this contour is manifestly self-conjugate under parity. We should point out that this contour differs from the one given in \cite{ABCTtwo} by only single minor appearing in $F_8^2$; however, this minor difference turns out to leave essentially all the geometry problems described in \cite{ABCTtwo} unchanged, and so the contour (\ref{g48_contour}) leads to precisely the same sum of twenty residues described in \cite{ABCTtwo}, and therefore reproduces the correct 8-point N$^2$MHV tree amplitude for all helicity configurations. 

As a further test of the validity of our contour prescription, let us briefly mention the tree-amplitude obtained for the $9$-point N$^2$MHV amplitude. As above, we may write,
\eq{\label{nine_pt_contour}
\hspace{-0.cm}\mathscr{A}_{9}^{(4)}=\frac{1}{\mathrm{vol[GL(}4\mathrm)]}\oint\limits_{\mathscr{F}_{9}^{(4)}=\vec{0}} \hspace{-0.15cm}\frac{dC_{\alpha\,a}\,\,\,\mathscr{H}_{9}^{(4)}}{(8)(1)(3)
\,\,\mathscr{F}_{9}^{(4)}}\,\,\prod_{\alpha=1}^4\delta^{4|4}\left(C_{\alpha\,a}\mathcal{W}_{a}\right),}
where $\mathscr{F}_{9}^{(4)}=F_7^1 F_7^2 \cdot F_8^1 F_8^2 \cdot F_9^1 F_9^2$ with each $F^j_{\ell}$ given explicitly by,
\eq{\label{9-point-n2mhv-contour}\hspace{-1.2cm}\mathscr{F}_9^{(4)}=\small\begin{tabular}{rrrcccll}
\multirow{2}{*}{$\left\{\rule[-17pt]{0pt}{35pt}\right.$}\!\!\!\!&$F_7^1=$&${\color{paper_red}(4)}\;\;\;\;\;{\color{black}(4\,7\,1\,2)}\;\;\;\,{\color{paper_red}(2)}\,\;\;\;{\color{paper_grey}(4\,6\,1\,3)}$&$-$&${\color{paper_grey}(4\,6\,7\,1)}{\color{paper_grey}(4\,1\,2\,3)}{\color{paper_grey}(4\,3\,5\,6)}{\color{paper_grey}(4\,7\,2\,5)}
$&$=\left[4\right]\bowtie S_{5\,6\,7\,1\,2\,3}$&\multirow{2}{*}{$\left.\rule[-17pt]{0pt}{35pt}\right\}$}\\
&$F_7^2=$&${\color{black}(1\,{\color{paper_blue}5\,6\,7})}{\color{black}({\color{paper_blue}1\,2\,3}\,7)}{\color{black}(1\,{\color{paper_blue}3\,4\,5})}{\color{paper_grey}(1\,2\,4\,6)}$&$-$&${\color{paper_grey}(1\,6\,7\,2)}{\color{paper_grey}(1\,2\,3\,4)}{\color{paper_grey}(1\,4\,5\,6)}{\color{paper_grey}(1\,7\,3\,5)}
$&$=\left[1\right]\bowtie S_{5\,6\,7\,2\,3\,4}$\\
\multirow{2}{*}{$\left\{\rule[-17pt]{0pt}{35pt}\right.$}\!\!\!\!&$F_8^1=$&$\;{\color{paper_red}(5)}\;\;\;\;\;{\color{black}(5\,8\,1\,2)}{\color{black}(5\,6\,2\,3)}{\color{paper_grey}(5\,7\,1\,3)}$&$-$&${\color{paper_grey}(5\,7\,8\,1)}{\color{paper_grey}(5\,1\,2\,3)}{\color{paper_grey}(5\,3\,6\,7)}{\color{paper_grey}(5\,8\,2\,6)}
$&$=\left[5\right]\bowtie S_{6\,7\,8\,1\,2\,3}$&\multirow{2}{*}{$\left.\rule[-17pt]{0pt}{35pt}\right\}$}\\
&$F_8^2=$&${\color{black}(1\,{\color{paper_blue}6\,7\,8})}{\color{black}({\color{paper_blue}1\,2\,3}\,8)}{\color{black}(1\,3\,4\,6)}{\color{paper_grey}(1\,2\,4\,7)}$&$-$&${\color{paper_grey}(1\,7\,8\,2)}{\color{paper_grey}(1\,2\,3\,4)}{\color{paper_grey}(1\,4\,6\,7)}{\color{paper_grey}(1\,8\,3\,6)}
$&$=\left[1\right]\bowtie S_{6\,7\,8\,2\,3\,4}$\\
\multirow{2}{*}{$\left\{\rule[-17pt]{0pt}{35pt}\right.$}\!\!\!\!&$F_9^1=$&$\;\;{\color{paper_red}(6)}\;\;\;\;\;{\color{black}(6\,{\color{paper_blue}9\,1\,2})}{\color{black}(6\,7\,2\,3)}{\color{paper_grey}(6\,8\,1\,3)}$&$-$&${\color{paper_grey}(6\,8\,9\,1)}{\color{paper_grey}(6\,1\,2\,3)}{\color{paper_grey}(6\,3\,7\,8)}{\color{paper_grey}(6\,9\,2\,7)}
$&$=\left[6\right]\bowtie S_{7\,8\,9\,1\,2\,3}$&\multirow{2}{*}{$\left.\rule[-17pt]{0pt}{35pt}\right\}$}\\
&$F_9^2=$&$\;\,{\color{paper_red}(7)}\,\;\,\;\;\;\;\;{\color{paper_red}(9)}\;\;\;\;\;{\color{black}(1\,3\,4\,7)}{\color{paper_grey}(1\,8\,2\,4)}$&$-$&${\color{paper_grey}(1\,8\,9\,2)}{\color{paper_grey}(1\,2\,3\,4)}{\color{paper_grey}(1\,4\,7\,8)}{\color{paper_grey}(1\,9\,3\,7)}$&$=\left[1\right]\bowtie S_{7\,8\,9\,2\,3\,4}$\end{tabular},}

Deforming this contour from the twistor string to $\mathcal{L}_{9,4}$ by sending each $t_{\ell}^j\to0$---removing all the contributions shown in coloured grey in (\ref{9-point-n2mhv-contour})---the problem of computing the tree-amplitude reduces to a series of `geometry problems'---finding the localization in the Grassmannian induced by requiring that each of the six maps $f_\ell^j$ vanish, and determining which of these configurations are supported entirely by the vanishing of consecutive minors.\footnote{Any configuration along the contour not entirely supported by consecutive minors will have vanishing residue because of the non-consecutive minors which constitute $\mathcal{H}_9^{(4)}$.} The six maps $f_\ell^j$ are given explicitly by,
\eq{\label{9-point-n2mhv-contour_fs}\hspace{-1.2cm}\mathcal{F}_9^{(4)}=\small\begin{tabular}{rrrcccll}
\multirow{2}{*}{$\left\{\rule[-17pt]{0pt}{35pt}\right.$}\!\!\!\!&$f_7^1=$&${\color{paper_red}(4)}\;\;\;\;\;{\color{black}(4\,7\,1\,2)}\;\;\;\,{\color{paper_red}(2)}\,\;\;\;$&\multirow{2}{*}{$\left.\rule[-17pt]{0pt}{35pt}\right\}$}\\
&$f_7^2=$&${\color{black}(1\,{\color{paper_blue}5\,6\,7})}{\color{black}({\color{paper_blue}1\,2\,3}\,7)}{\color{black}(1\,{\color{paper_blue}3\,4\,5})}$\end{tabular}\bigcup\begin{tabular}{rrrcccll}
\multirow{2}{*}{$\left\{\rule[-17pt]{0pt}{35pt}\right.$}\!\!\!\!&$f_8^1=$&$\;{\color{paper_red}(5)}\;\;\;\;\;{\color{black}(5\,8\,1\,2)}{\color{black}(5\,6\,2\,3)}$&\multirow{2}{*}{$\left.\rule[-17pt]{0pt}{35pt}\right\}$}\\
&$f_8^2=$&${\color{black}(1\,{\color{paper_blue}6\,7\,8})}{\color{black}({\color{paper_blue}1\,2\,3}\,8)}{\color{black}(1\,3\,4\,6)}$\end{tabular}\bigcup\begin{tabular}{rrrcccll}
\multirow{2}{*}{$\left\{\rule[-17pt]{0pt}{35pt}\right.$}\!\!\!\!&$f_9^1=$&$\;\;{\color{paper_red}(6)}\;\;\;\;\;{\color{black}(6\,{\color{paper_blue}9\,1\,2})}{\color{black}(6\,7\,2\,3)}$&\multirow{2}{*}{$\left.\rule[-17pt]{0pt}{35pt}\right\}$}\\
&$f_9^2=$&$\;\,{\color{paper_red}(7)}\,\;\,\;\;\;\;\;{\color{paper_red}(9)}\;\;\;\;\;{\color{black}(1\,3\,4\,7)}$\end{tabular}.}
We have found that there are 50 non-vanishing, consecutively-supported residues along the contour (\ref{9-point-n2mhv-contour}) and that these residues perfectly reproduce the fully-supersymmetric $9$-point N$^2$MHV tree amplitude.

These 50 terms, together with the `geometry problems' giving rise to each, are collected in appendix \ref{9ptappendix}, where we have followed the conventions of \cite{ABCTtwo} for the naming of each residue according to its localization in $C_{\alpha\,a}$.

\subsection{N$^3$MHV Amplitudes and Beyond}
As was the case for the $7$-point amplitude, the parity conjugate of the $9$-point N$^2$NHV amplitude represents the first sufficiently-general N$^3$MHV amplitude from which we may `bootstrap' the general N$^3$MHV result. We will see that by requiring the $9$-point N$^3$MHV amplitude to be iteratively-related to the $8$-point N$^3$MHV amplitude---itself obtained as the parity-conjugate of the $8$-point NMHV amplitude---will uniquely fix the structure of the ansatz for all further amplitudes in $\mathcal{N}=4$ super Yang-Mills.

Taking the parity conjugate of the $9$-point $k=4$ contour (\ref{9-point-n2mhv-contour}), we find,
\eq{\hspace{-1cm}\mathscr{F}_9^{(5)}=\widetilde{\mathscr{F}_9^{(4)}}=\left\{\begin{array}{lr}F_8^1=&\left[4\,5\right]\bowtie S_{6\,7\,8\,1\,2\,3}\\F_8^2=&\left[1\,5\right]\bowtie S_{6\,7\,8\,2\,3\,4}\\F_8^3=&\left[1\,2\right]\bowtie S_{6\,7\,8\,3\,4\,5}\\\end{array}\right\}\bigcup\left\{\begin{array}{lr}F_9^1=&\left[5\,6\right]\bowtie S_{7\,8\,9\,1\,2\,3}\\F_9^2=&\left[1\,6\right]\bowtie S_{7\,8\,9\,2\,3\,4}\\F_9^3=&\left[1\,2\right]\bowtie S_{7\,8\,9\,3\,4\,5}\\\end{array}\right\}.} 
Notice that only the last three $F_\ell^j$'s---those of the second set above---involve column $9$. Moreover, all of the $F_\ell^j$'s for $\ell=8$ involve column $8$. Therefore, the requirement that the $9$-point N$^3$MHV contour is the extension of the $8$-point N$^3$MHV contour, uniquely fixes the $\ell$-dependence of the maps $F_{\ell}^j$. With this, it is not hard to see that the general solution for all N$^3$MHV amplitudes is given by 
\eq{\mathscr{F}_n^{(5)}=\prod_{\ell=8}^n\Big(F_{\ell}^1\cdot F_{\ell}^2\cdot F_{\ell}^3\Big),\quad\mathrm{with}\quad\left\{\begin{array}{lr}F_\ell^1=&\left[\ell\smallminus4\,\ell\smallminus3\right]\bowtie S_{\ell-2\,\,\ell-1\,\,\ell\,\,1\,\,2\,\,3}\\F_\ell^2=&\left[1\,\ell\smallminus3\right]\bowtie S_{\ell-2\,\,\ell-1\,\,\ell\,\,2\,\,3\,\,4}\\F_\ell^3=&\left[1\,\,2\right]\bowtie S_{\ell-2\,\,\ell-1\,\,\ell\,\,3\,\,4\,\,5}\end{array}\right\}.} 

As a concrete illustration of this contour, consider for example the 10-point N$^3$MHV amplitude,
\eq{
\hspace{-0.cm}\mathscr{A}_{10}^{(5)}=\frac{1}{\mathrm{vol[GL(}5\mathrm)]}\oint\limits_{\mathscr{F}_{10}^{(5)}=\vec{0}} \hspace{-0.15cm}\frac{dC_{\alpha\,a}\,\,\,\mathscr{H}_{10}^{(5)}}{(9)(1)(3)
\,\,\mathscr{F}_{10}^{(5)}}\,\,\prod_{\alpha=1}^5\delta^{4|4}\left(C_{\alpha\,a}\mathcal{W}_{a}\right),
}\definecolor{ggreen}{rgb}{0.05,0.45,0.125}
where $\mathscr{F}_{10}^{(5)}=F_8^1 F_8^2 F_8^3 \cdot F_9^1 F_9^2 F_9^3 \cdot F_{10}^1 F_{10}^2 F_{10}^3$, and with each $F_\ell^j$ given by
\eq{\hspace{-1.6cm}\mathscr{F}_{10}^{(5)}=\small
\begin{tabular}{r@{}rrccc@{}l@{}l@{}}\multirow{3}{*}{$\left\{\rule[-30pt]{0pt}{-15pt}\right.$}
&$F_8^1=$&$\;\;\;{\color{paper_red}(4)}\;\;\;\,\;\;\;{\color{black}(4\,5\,8\,1\,2)}\;\;\;\;\;{\color{paper_red}(2)}\;\;\;\;\;{\color{paper_grey}(4\,5\,7\,1\,3)}$&$-$&${\color{paper_grey}(4\,5\,1\,2\,3)}{\color{paper_grey}(4\,5\,3\,6\,7)}{\color{paper_grey}(4\,5\,7\,8\,1)}{\color{paper_grey}(4\,5\,2\,6\,8)}$&$=\left[4\,5\right]\bowtie S_{6\,7\,8\,\,1\,2\,3}$&\multirow{3}{*}{$\left.\rule[-30pt]{0pt}{-15pt}\right\}$}\\
&$F_8^2=$&${\color{black}(1\,{\color{paper_blue}5\,6\,7\,8})}{\color{black}({\color{ggreen}1\,2\,3}\,5\,8)}{\color{black}(1\,{\color{paper_blue}3\,4\,5\,6})}{\color{paper_grey}(1\,5\,7\,2\,4)}$&$-$&${\color{paper_grey}(1\,5\,2\,3\,4)}{\color{paper_grey}(1\,5\,4\,6\,7)}{\color{paper_grey}(1\,5\,7\,8\,2)}{\color{paper_grey}(1\,5\,3\,6\,8)}$&$=\left[1\,5\right]\bowtie S_{6\,7\,8\,\,2\,3\,4}$\\&$F_8^3=$&${\color{black}(1\,2\,{\color{ggreen}6\,7\,8})}{\color{black}({\color{paper_blue}1\,2\,3\,4}\,8)}{\color{black}(1\,2\,{\color{ggreen}4\,5\,6})}{\color{paper_grey}(1\,2\,7\,3\,5)}$&$-$&${\color{paper_grey}(1\,2\,3\,4\,5)}{\color{paper_grey}(1\,2\,5\,6\,7)}{\color{paper_grey}(1\,2\,7\,8\,3)}{\color{paper_grey}(1\,2\,4\,6\,8)}$&$=\left[1\,2\right]\bowtie S_{6\,7\,8\,3\,4\,5}$&\\\multirow{3}{*}{$\left\{\rule[-30pt]{0pt}{-15pt}\right.$}
&$F_9^1=$&$\;\;\;{\color{paper_red}(5)}\;\;\;\;\;\;\;{\color{black}(5\,6\,9\,1\,2)}{\color{black}(2\,3\,{\color{ggreen}5\,6\,7})}{\color{paper_grey}(5\,6\,8\,1\,3)}$&$-$&${\color{paper_grey}(5\,6\,1\,2\,3)}{\color{paper_grey}(5\,6\,3\,7\,8)}{\color{paper_grey}(5\,6\,8\,9\,1)}{\color{paper_grey}(5\,6\,2\,7\,9)}$&$=\left[5\,6\right]\bowtie S_{7\,8\,9\,\,1\,2\,3}$&\multirow{3}{*}{$\left.\rule[-30pt]{0pt}{-15pt}\right\}$}\\
&$F_9^2=$&${\color{black}(1\,{\color{paper_blue}6\,7\,8\,9})}{\color{black}({\color{ggreen}1\,2\,3}\,6\,9)}{\color{black}(1\,3\,4\,6\,7)}{\color{paper_grey}(1\,6\,8\,2\,4)}$&$-$&${\color{paper_grey}(1\,6\,2\,3\,4)}{\color{paper_grey}(1\,6\,4\,7\,8)}{\color{paper_grey}(1\,6\,8\,9\,2)}{\color{paper_grey}(1\,6\,3\,7\,9)}$&$=\left[1\,6\right]\bowtie S_{7\,8\,9\,\,2\,3\,4}$\\&$F_9^3=$&${\color{black}(1\,2\,{\color{ggreen}7\,8\,9})}{\color{black}({\color{paper_blue}1\,2\,3\,4}\,9)}{\color{black}(1\,2\,4\,5\,7)}{\color{paper_grey}(1\,2\,8\,3\,5)}$&$-$&${\color{paper_grey}(1\,2\,3\,4\,5)}{\color{paper_grey}(1\,2\,5\,7\,8)}{\color{paper_grey}(1\,2\,8\,9\,3)}{\color{paper_grey}(1\,2\,4\,7\,9)}$&$=\left[1\,2\right]\bowtie S_{7\,8\,9\,\,3\,4\,5}$&\\\multirow{3}{*}{$\left\{\rule[-30pt]{0pt}{-15pt}\right.$}
&$F_{10}^1=$&$\;\;\;{\color{paper_red}(6)}\,\;\;\;\;\;{\color{black}(6\,7\,{\color{ggreen}10\,1\,2})}{\color{black}(2\,3\,{\color{ggreen}6\,7\,8})}{\color{paper_grey}(6\,7\,9\,1\,3)}$&$-$&${\color{paper_grey}(6\,7\,1\,2\,3)}{\color{paper_grey}(6\,7\,3\,8\,9)}{\color{paper_grey}(6\,7\,9\,10\,1)}{\color{paper_grey}(6\,7\,2\,8\,10)}$&$=\left[6\,7\right]\bowtie S_{8\,9\,10\,1\,2\,3}$&\multirow{3}{*}{$\left.\rule[-30pt]{0pt}{-15pt}\right\}$}\\
&$F_{10}^2=$&$\;\;\;{\color{paper_red}(7)}\;\;\;\;\,\;{\color{black}(7\,{\color{paper_blue}10\,1\,2\,3})}{\color{black}(1\,3\,4\,7\,8)}{\color{paper_grey}(1\,7\,9\,2\,4)}$&$-$&${\color{paper_grey}(1\,7\,2\,3\,4)}{\color{paper_grey}(1\,7\,4\,8\,9)}{\color{paper_grey}(1\,7\,9\,10\,2)}{\color{paper_grey}(1\,7\,3\,8\,10)}$&$=\left[1\,7\right]\bowtie S_{8\,9\,10\,2\,3\,4}$\\
&$F_{10}^3=$&$\;\;\;{\color{paper_red}(8)}\;\;\;\;\;\;\;\;\;\;{\color{paper_red}(10)}\;\;\;\;\;{\color{black}(1\,2\,4\,5\,8)}{\color{paper_grey}(1\,2\,9\,3\,5)}$&$-$&${\color{paper_grey}(1\,2\,3\,4\,5)}{\color{paper_grey}(1\,2\,5\,8\,9)}{\color{paper_grey}(1\,2\,9\,10\,3)}{\color{paper_grey}(1\,2\,4\,8\,10)}$&$=\left[1\,2\right]\bowtie S_{8\,9\,10\,3\,4\,5}$\end{tabular}}

\noindent and again $\mathscr{{H}}_{10}^{(5)}$ can be simply read-off from $F^j_{\ell}$'s 
$${\begin{split}\mathscr{{H}}_{10}^{(5)}=&\phantom{\,\times\,}{\color{black}(4\,5\,8\,1\,2)}{\color{black}(1\,5\,6\,7\,8)}{\color{black}(1\,5\,8\,2\,3)}{\color{black}(1\,5\,3\,4\,6)}{\color{black}(1\,2\,6\,7\,8)}{\color{black}(1\,2\,8\,3\,4)}{\color{black}(1\,2\,4\,5\,6)}\\ & \times {\color{black}(5\,6\,9\,1\,2)} {\color{black}(5\,6\,2\,3\,7)}{\color{black}(1\,6\,7\,8\,9)}{\color{black}(1\,6\,9\,2\,3)}{\color{black}(1\,6\,3\,4\,7)}{\color{black}(1\,2\,7\,8\,9)}{\color{black}(1\,2\,9\,3\,4)}\\ & \times {\color{black}(1\,2\,4\,5\,7)}{\color{black}(6\,7\,10\,1\,2)}{\color{black}(6\,7\,2\,3\,8)}{\color{black}(1\,7\,10\,2\,3)}{\color{black}(1\,7\,3\,4\,8)}{\color{black}(1\,2\,4\,5\,8)}\\ & \times {\color{paper_grey}(4\,5\,7\,1\,3)}{\color{paper_grey}(1\,5\,7\,2\,4)}{\color{paper_grey}(1\,2\,7\,3\,5)}{\color{paper_grey}(5\,6\,8\,1\,3)}{\color{paper_grey}(1\,6\,8\,2\,4)}{\color{paper_grey}(1\,2\,8\,3\,5)} {\color{paper_grey}(6\,7\,9\,1\,3)}{\color{paper_grey}(1\,7\,9\,2\,4)}{\color{paper_grey}(1\,2\,9\,3\,5)}  \, .\end{split}}$$

Although it would require more space than warranted by an appendix, we have explicitly verified that the contour above includes 175 non-vanishing residues which precisely matches the general, $10$-point N$^3$MHV amplitude. 

Continuing in this manner, we arrive at the general formula (\ref{general_formula}),
\eq{\nonumber\hspace{-0.cm}\mathscr{A}_{n}^{(k)}=\frac{1}{\mathrm{vol[GL(}k\mathrm)]}\oint\limits_{\mathscr{F}_n^{(k)}=\vec{0}} \hspace{-0.15cm}\frac{dC_{\alpha\,a}\,\,\,\mathscr{H}_{n}^{(k)}}{(n-1)(1)(3)
\,\,\mathscr{F}_{n}^{(k)}}\,\,\prod_{\alpha=1}^k\delta^{4|4}\left(C_{\alpha\,a}\mathcal{W}_{a}\right),
}
where $\mathscr{F}_n^{(k)}=(F_{k+3}^1\cdots F_{k+3}^{k-2})\cdot(F_{k+4}^1\cdots F_{k+4}^{k-2})\cdots(F_n^1\cdots F_n^{k-2})$ with each $F_\ell^j$ given by 
\eq{F_\ell^j\equiv\sigma_\ell^j\bowtie S_{\ell-2\,\,\ell-1\,\,\ell\,\,j\,\,j+1\,\,j+2}.}

\newpage
\subsection{General Properties of the Result}

\subsubsection{Parity}

One of the important features of the general contour obtained in the previous subsections is that it is manifestly parity-symmetric. By this, we mean that the parity-conjugate of a given amplitude's contour is the contour for the parity-conjugate amplitude. For example, for all $n=2k$, the contour given by $\mathscr{F}_{n=2k}^{(k)}$ is manifestly parity self-conjugate. 

To see how this works, consider the role played by each of the $n$ columns of the Grassmannian $C_{\alpha\,a}$ in the definition of the Veronese map $F_\ell^j\equiv\flist\bowtie S_{\ell-2\,\ell-1\,\ell\,j\,j+1\,j+2}$. In general, the $n$ columns break into six contiguous groups, 
\eq{{\color{paper_blue}\underbrace{[1\,2\,\,\,\cdots\,\,\,j\smallminus1]}_{\phantom{\flist_{\ell}^j}\in \flist_{\ell}^j}}\,\,\,{\color{paper_red}\overbrace{[j\,\,j\smallplus1\,\,j\smallplus2]}^{\phantom{S_{\ell}^j}\in S_{\phantom{\ell}}^{\phantom{j}}}}\,\,\,{\color{paper_grey} [j\smallplus3\,\,\,\cdots\,\,\,j\smallplus(k\smallminus\ell)\smallminus1]}\,\,\,{\color{paper_blue} \underbrace{[j\smallplus(k\smallminus\ell)\,\,\,\cdots\,\,\,\ell\smallminus3]}_{\phantom{\flist_{\ell}^j}\in \flist_{\ell}^j}}\,\,\,{\color{paper_red} \overbrace{[\ell\smallminus2\,\,\ell\smallminus1\,\,\ell]}^{\phantom{S_{\ell}^j}\in S_{\phantom{\ell}}^{\phantom{j}}}}\,\,\,{\color{paper_grey} [\ell\smallplus1\,\,\,\cdots\,\,\,n]},\nonumber}
where the columns of $C_{\alpha\,a}$ which do not participate at all in $F_\ell^j$ have been coloured grey to emphasize the `gaps' in the roles played by various columns. Importantly, parity does not change the contiguousness of these six groups, or the roles they played by the six columns of the primative Veronese map $S_{\ell-2\,\ell-1\,\ell\,j\,j+1\,j+2}$---coloured red above; parity merely changes the labels we assign each column, and exchanges the $k-6$ columns involved in {\it all} the minors of $F_\ell^j$---those of $\flist_\ell^j$, coloured blue above---with the $n-k-6$ columns involved in {\it none} of the minors of $F_\ell^j$---those coloured grey above. That is,
\eq{
\left\{\begin{array}{lcl}{\color{paper_blue}[1\,\,\,\cdots\,\,\,j\smallminus1]}&\multirow{6}{*}{$\left.\rule[-60pt]{0pt}{-25pt}\right\}\xrightarrow
[\substack{k\mapsto(n-k)\\i\mapsto
(n+1)-i}]{\text{parity}}{}\left\{\rule[-60pt]{0pt}{-25pt}\right.$}&\qquad\qquad\qquad\qquad\quad{\color{paper_grey} [n\smallminus j\smallplus2\,\,\cdots\,\,n]}\\
\quad{\color{paper_red}[j\,\,j\smallplus1\,\,j\smallplus2]}&&\qquad\qquad\qquad{\color{paper_red}[n\smallminus j\smallminus1\,\,n\smallminus j\,\,n\smallminus j\smallplus1]}\\
\qquad{\color{paper_grey} [j\smallplus3\,\cdots\,j\smallplus(k\smallminus\ell)\smallminus1]}&&\qquad\qquad{\color{paper_blue}[n\smallplus\ell\smallminus j\smallminus k\,\cdots\,n\smallminus j\smallminus2]}\\
\quad\qquad\qquad{\color{paper_blue}[j\smallplus(k\smallminus\ell)\,\cdots\,\ell\smallminus3]}&&\qquad{\color{paper_grey} [n\smallminus\ell\smallplus4\,\cdots\,n\smallplus\ell\smallminus j\smallminus k\smallminus1]}\\
\qquad\quad\qquad\qquad\qquad{\color{paper_red}[\ell\smallminus2\,\,\ell\smallminus1\,\,\ell]}&&\quad{\color{paper_red}[n\smallminus\ell\smallplus1\,\,\,n\smallminus\ell\smallplus2\,\,\,n\smallminus\ell\smallplus3]}\\
\qquad\qquad\qquad\qquad\qquad{\color{paper_grey} [\ell\smallplus1\,\,\,\cdots\,\,\,n]}&&{\color{paper_blue}[1\,\,\,\cdots\,\,\,n\smallminus\ell]}
\end{array}\right\}.}
This shows that,
\eq{F_\ell^j\xrightarrow[\substack{k\mapsto(n-k)\\i\mapsto
(n+1)-i}]{\text{parity}}{}\widetilde{F_\ell^j}=F_{(n-j+1)}^{(n-\ell+1)}\equiv
F_{\ell'}^{j'},\vspace{-0.4cm}} so that
\eq{\label{parity_of_F}\hspace{-1.2cm}\mathscr{F}_n^{(k)}=\prod_{\ell=k+3}^n\left(\prod_{j=1}^{k-2}F_{\ell}^j\right)\xrightarrow
[\substack{k\mapsto(n-k)\\i\mapsto
(n+1)-i}]{\text{parity}}{}{\mathscr{F}}_{n}^{(n-k)}=\prod_{\ell=k+3}^{n}\left(\prod_{j=1}^{k-2}
\widetilde{F_{\ell}^j}\right)=\prod_{j'=1}^{k'-2}\left(\prod_{\ell'=k'+3}^{n}F_{\ell'}^{j'}\right)=
\prod_{\ell'=k'+3}^n\left(\prod_{j'=1}^{k'-2}F_{\ell'}^{j'}\right),}
where $k'\equiv (n-k)$, being what it was required to demonstrate.

\subsubsection{Manifest Soft-Limits and the Particle Interpretation}
As we have seen, the contour integral giving the $n-1$-particle N$^{(k-2)}$MHV scattering amplitude, is related to that giving the $n$-particle N$^{(k-2)}$MHV scattering amplitude by a single overall factor which relates $\mathscr{H}_n^{(k)}$ to $\mathscr{H}_{n-1}^{(k)}$, together with a partial contour specification, 
\begin{align}
\hspace{-1cm}\nonumber\mathscr{A}_{n}^{(k)}=&\frac{1}{\mathrm{vol[GL(}k\mathrm{)]}}\oint\limits_{\mathscr{F}_{n}^{(k)}=\vec{0}}\!\!\!\! dC_{\alpha\,a}\frac{\mathscr{H}_n^{(k)}}{(n-1)(1)(3)\,\,\mathscr{F}_{n}^{(k)}}\\=&\frac{1}{\mathrm{vol[GL(}k\mathrm{)]}}\oint\limits_{\mathscr{F}_{n-1}^{(k)}=\vec{0}}\!\!\!\! dC_{\alpha\,\hat{a}}\frac{\mathscr{H}_{n-1}^{(k)}}{(1)(3)\,\,\mathscr{F}_{n-1}^{(k)}}\times\oint\limits_{\substack{F_{n}^1=0\\[-0.1cm]\vdots\\F_n^{k-2}=0}}\!\!\!\!d C_{\alpha n}\frac{\mathscr{H}_n^{(k)}/\mathscr{H}_{n-1}^{(k)}}{(n-1)\,\,F_{n}^{1}\cdot F_{n}^{2}\cdots F_{n}^{k-2}},\label{n_to_n-1}
\end{align}
where $\hat{a}=1,\ldots,n-1$ and the ratio $\mathscr{H}_{n}^{(k)}/\mathscr{H}_{n-1}^{(k)}$ was given explicitly after equation (\ref{explicit_veronese_maps}) in section 2. This separation of the integral is warranted because only the maps $F_n^{1},\ldots, F_n^{k-2}$ involve the variables of the $n^{\mathrm{th}}$ column of $C_{\alpha\,a}$. We can anticipate which contour should be specified for these $k-2$ variables to extract the soft-limit by considering the duality between the geometry of the columns of $C_{\alpha\,a}$, viewed as points in $\mathbb{P}^{k-1}$, and $\mathcal{Z}$-twistor-space geometry \cite{ABCTtwo}. In twistor space, the soft-limit is achieved when the three twistors $\mathcal{Z}_{n-1},\mathcal{Z}_n,$ and $\mathcal{Z}_1$ become (projectively) collinear, and so we can extract the soft limit from $\mathscr{A}_n^{(k)}$ by choosing a contour for which the column-vectors $C_{\alpha\,n-1},C_{\alpha\,n},$ and $C_{\alpha\,1}$ become linearly-dependent. This fixes exactly $(k-2)$ variables of integration, and so should completely specify the integral factor in (\ref{n_to_n-1}) relating $\mathscr{A}_{n}^{(k)}$ to $\mathscr{A}_{n-1}^{(k)}$. 

Recalling the definition of the maps $F_n^1,F_n^2,\ldots F_n^{k-1}$, it is easy to see that when the columns $n-1,n,1$ become linearly-dependent, $F_n^2,\ldots,F_n^{k-2}$ all vanish, while $F_{n}^{1}$ factorizes into simply the product of four minors. Importantly, notice that $\mathscr{H}_n^{(k)},\mathscr{H}_{n-1}^{(k)},$ and all the factors of $\mathscr{F}_{n-1}^{(k)}$ are regular in this limit. Because of this, we can apply the global residue theorem in (\ref{n_to_n-1}) to trade $F_n^1$ for the minor $(n-1)$---which does vanish in this limit. 

This allows us to view the contour integral for the twistor string entirely in $\mathcal{L}_{n,k}$, and refer to some well-known facts \cite{ABCCKT:2010,ABCTtwo} relating residues in $\mathcal{L}_{n,k}$ to those of $\mathcal{L}_{n-1,k}$ to see how the soft-factor arises. It turns out that the contour which sets three consecutive columns of the Grassmannian to be linearly dependent is particularly nice, and is nothing but a holomorphic inverse soft-factor times the ratio of the $k$ consecutive minors containing $n$ to the $k-1$ minors which were consecutive only prior to `adding particle $n$' to $G(k,n-1)$. Recall that this ratio of minors is precisely built-into the definition of $\mathscr{H}_n^{(k)}$

\section{Transformation to the Twistor String in Link Variables}

In this section we demonstrate the equivalence of the
twistor string amplitude \cite{RSV} (when expressed in link variables
as in \cite{SV,DG}) to our main formula (\ref{general_formula}) above.
This is accomplished via repeated application of the identity
transformation
\begin{equation}
\begin{aligned} \label{identity}
\delta(S_{ijkrst}) \delta(S_{ijkrsu}) \sim
\frac{(jkt)(irt)}{(jks)(irs)}
\delta(S_{ijkrst}) \delta(S_{ijkrtu });
\end{aligned}
\end{equation} 
here, $\sim$ is used to indicate that the replacement may be made
at the level of the integrand only strictly for {\it physical configurations} along the contour of integration. This transformation played an important role in the analysis of \cite{NVW}, \cite{DGtwo}. Note that this relation indicates a specific change in the
contour prescription: the
$\delta(S_{ijkrsu})$ on the left-hand
side may localize the integral on fewer (or more) poles than
the $\delta(S_{ijkrtu})$ on the right, in which case the extra (or missing)
poles on the right-hand side are provided by zeros the minors in
the denominator (or cancelled by zeros of the minors in the numerator).

In the next two subsections we first focus on following the
transformation of the $\delta(F^i_{\ell})$'s from (\ref{general_formula}) to the formula $(4.12)$ in \cite{DG}.  We then collect all the pre-factors
which pile-up along the way and demonstrate precise agreement
with \cite{DG}.
It is very easy to check the agreement between our formula
and that of \cite{DG} for NMHV using \cite{NVW}.
We may proceed by induction at step $n$,
beginning
with the assumption that (\ref{general_formula}) agrees with \cite{DG} for $n{-}1$-points.

\subsection{Transforming the $\delta(F_\ell^j)$'s}

Let us first transform the $\delta(F_\ell^j)$'s from (\ref{general_formula}) to those in \cite{DG}.
Because we will use induction, we only need to consider $F_n^j$
which we will denote $F_j.$
In order to compare with \cite{DG} we must
first change the common piece in $F_{j}$,
namely $\sigma^{j}_n=[1,2,\ldots,j-1,n+j-k,\ldots,n-3]$ in (\ref{F}), into a subset of the columns
$[1,2,\ldots,k]$.\footnote{The meaning of this will become clear by looking at the final result, equation (\ref{newsextics}).}
In this sense $F_1$ is the `worst' of the $F$'s and $F_{k-2}$ is the
`best', so the strategy will be to first make all transformations on
$F_1$, then to make all transformations on $F_2$, and continue
in the same way (as far as possible) until $F_{k-3}$.  In this
way we gradually transform all of the original $\delta(F_j)'s$ into
`real sextics' (objects which are indeed sextics in a certain gauge).  In the following we show a first few steps
and then move on to the final conclusion.

$\bullet$ Let us first show how to transform $F_1$ to $F''_1,$ 
\begin{equation}
\begin{aligned}
F_1&=[n-k+1~\cdots~n-3] \bowtie S_{1~2~3~n-2~n-1~n}\cr
 \rightarrow F''_1&=[n-k+2~\cdots~n-3~2]\bowtie S_{1~3~4~n-2~n-1~n}.
\label{s1pp}
\end{aligned}
\end{equation}

Step one is to use the identity
\begin{equation}
\begin{aligned} \label{tranform1}
\delta(F_1) \delta (F'_2) \sim J^{(1)}_1 \delta(F'_1) \delta (F'_2),
\end{aligned}
\end{equation}
where the sextics and the Jacobian are
\begin{equation}
\begin{aligned}
F'_{2}=&[n-k+2~\cdots~n-3~1] \bowtie S_{2~3~n-k+1~n-2~n-1~n}\\
F'_{1}=&[n-k+2~\cdots~n-3~2] \bowtie S_{1~3~n-k+1~n-2~n-1~n}\\
J^{(1)}_1=&[n-k+2~\cdots~n-3] \bowtie {(n~1~2~3)(n-2~n-1~1~2)\over (n~1~3~n-k+1)(n-2~n-1~1~n-k+1)}.
\end{aligned}
\end{equation}
This identity follows from (\ref{identity}) by setting a particular gauge, namely to use GL$(k)$ symmetry to set $k$ columns $[1,2,3,n-k+1 \cdots n-3]$ of $k \times n$ matrix $(C_{\alpha \,a})$ to be an identity square matrix, and we will denote the gauge as $\{1,2,3,n-k+1 \cdots n-3 \}$. Note that we also transformed $F_2$ into $F'_2$ which generated a Jacobian $J$
which will end up canceling, so we will not  write it explicitly. 

Next we further transform $F'_1$ by using
\begin{equation}
\begin{aligned}
\delta(F'_1) \delta (F^{(n-1)}_1) \sim J^{(2)}_1 \delta(F''_1) \delta (F^{(n-1)}_1),
\end{aligned}
\end{equation}
where
\begin{equation}
\begin{aligned}
F^{(n-1)}_{1}=&[n-k+2~\cdots~n-3~2] \bowtie S_{1~3~n-k+1~n-2~n-1~4},\\
F''_{1}=&[n-k+2~\cdots~n-3~2] \bowtie S_{1~3~4~n-2~n-1~n},\\
J^{(2)}_{1_j}=&[n-k+2~\cdots~n-3~\bar{j}] \bowtie {(n-1~4~j)(3~n-2~4)\over (n-1~n-k+1~j)(3~n-2~n-k+1)},
\end{aligned}
\end{equation}
with $j=1$ and $\bar{j}=2$.
Note that in carrying out these transformation we have
made use of the constraint $F_1^{(n-1)}$ 
which can be obtained by transforming $F^j_{n-1}$ of $n-1$ point.

The third step is to transform $F_2'$ back to $F_2$, which
generates Jacobian $J^{-1}$. 

To summarize the construction so far, we have shown how to
transform
the original $F_1$ into a ``better" quantity $F''_1$ 
at the cost of inserting the Jacobain factor
$J^{(1)}_1 J^{(2)}_{1_1}$ into the integrand.

$\bullet$ Next we would like to similarly process $F_{2}$ with $F''_{1}$.  By applying (\ref{identity})
for the new $F''_{1}$ and the old $F_{2}$
\begin{equation}
\begin{aligned}
F''_{1}=&[n-k+2~\cdots~n-3~2]\bowtie S_{1~3~4~n-2~n-1~n},\\
F_{2}=&[n-k+2~\cdots~n-3~1]\bowtie S_{2~3~4~n-2~n-1~n},
\end{aligned}
\end{equation}
we get the new quantities
\begin{equation}
\begin{aligned}
F'''_{1}=&[n-k+3~\cdots~n-3~2~3]\bowtie S_{1~4~5~n-2~n-1~n},\\
F'''_{2}=&[n-k+3~\cdots~n-3~1~3]\bowtie S_{2~4~5~n-2~n-1~n}.
\end{aligned}
\end{equation}
The Jacobians generated from this step are
\begin{equation}
J^{(1)}_{2} J^{(2)}_{2_1} J^{(1)}_{2} J^{(2)}_{2_2},
\end{equation}
where
\begin{equation}
\begin{aligned}
J^{(1)}_{2}&=[n-k+3~\cdots~n-3] \bowtie {(n~1~2~3~4)(n-2~n-1~1~2~3)\over (n~1~2~4~n-k+2)(n-2~n-1~1~2~n-k+2)},\\
J^{(2)}_{2_j}&=[n-k+3~\cdots~n-3~\bar{j}] \bowtie {(n-1~5~j)(4~n-2~5)\over (n-1~n-k+2~j)(4~n-2~n-k+2)},
\end{aligned}
\end{equation}
where $j=1, 2$ and $\bar{1}=(2, 3)$, $\bar{2}=(1, 3)$. 

$\bullet$ We proceed
by transforming the original $F_3$ together with the new
$F_1''', F_2'''$ into three new quantities $F_1''''$, $F_2''''$,
$F_3''''$.
We continue in this manner until we reach $F^{'''\cdots''}_{k-3}$. 
In each step we will always  make two-type transformations like the ones described above. 
At the end of the day, we have new quantities
\begin{equation}
\begin{aligned} \label{newsextics}
F_{j}=[1,2,\cdots, \cancel{j}, \cdots, k-2] \bowtie S_{j~k-1~k~n-2~n-1~n},
\end{aligned}
\end{equation}
where $1 \le j \le k-2$. 
The Jacobians generated during the whole process are products of
\begin{equation}
\begin{aligned}
J^{(1)}_{l}&=[n\smallminus k\smallplus \ell\smallplus 1~\cdots~n\smallminus 3] \bowtie {(n~1~2~\cdots~\ell\smallplus 2)(n\smallminus 2~n\smallminus 1~1~\cdots~\ell\smallplus 1)\over (n~1~\cdots~\ell~\ell\smallplus 2~n\smallminus k\smallplus \ell)(n\smallminus 2~n\smallminus1~1~\cdots~\ell~~n\smallminus k\smallplus \ell)},\\
J^{(2)}_{\ell_j}&=[n\smallminus k\smallplus \ell\smallplus 1~\cdots~n\smallminus 3~\bar{j}] \bowtie {(n\smallminus 1~\ell\smallplus 3~j)(\ell\smallplus 2~n\smallminus 2~\ell\smallplus 3)\over (n\smallminus 1~n\smallminus k\smallplus \ell~j)(\ell\smallplus 2~n\smallminus 2~n\smallminus k\smallplus \ell)},
\end{aligned}
\end{equation}
where $\bar{j}=(1,2,\cdots, \cancel{j}, \cdots, \ell+1)$, $1 \le \ell \le k-3$ and $1 \le j \le \ell$. 

Finally let us choose a gauge $\{1,2,3, \cdots, k\}$, in
which case $F_j=S_{j~k-1~k~n-2~n-1~n}$ may be found in (\ref{newsextics}).
Thus we have mapped our $F^j_{n}$'s to the sextics in \cite{DG},
and all we are left to compare is the corresponding prefactor.

\subsection{Collecting Prefactors}

Let us now verify that performing the above procedure on our
formula (\ref{general_formula})leads to precisely the same prefactor inside
the integral as in \cite{DG}.
We only need to compare the ratio between $n$-point amplitude and $(n-1)$-point amplitude
which for our formula (\ref{general_formula}) reads
\begin{equation}
\begin{aligned}
A_n={\mathscr{H}_{n}^{(k)} \over \mathscr{H}^{(k)}_{n- 1}}=&\phantom{\,\times\,}{1 \over (n \smallminus 1~n~1~\cdots~k\smallminus 2)} \big[\prod^{k\smallminus 1}_{j=1} (n \smallminus k\smallplus j~\cdots~n\smallminus1~1~\cdots~j)
\\ &\times (n\smallminus k\smallplus j~\cdots~n\smallminus 3~n~1~\cdots~j\smallplus 1)
 (n\smallminus k\smallplus j~\cdots~n\smallminus 2~1~\cdots~j\smallminus 1~j\smallplus 1~j\smallplus 2)\\  
 &\times (n\smallminus k\smallplus i~\cdots ~ n\smallminus 3~n\smallminus 1~1~\cdots~j~j\smallplus 2)\big].
       \end{aligned}
\end{equation}
The corresponding ratio in twistor string is given by the formula $(4.12)$ of \cite{DG} .
Taking into account all the Jacobians from the transformations
described in the previous subsection, we find the ratio 
of our formula (\ref{general_formula}) to that in \cite{DG} is
equal to one.
This completes the proof.

\section*{Acknowledgments}

We are grateful to N.~Arkani-Hamed, F.~Cachazo, R.~Roiban, D.~Skinner, 
M.~Spradlin and C.~Vergu for very helpful conversations.  The work of AV and CW was
supported in part by the US Department of Energy under contract
DE-FG02-91ER40688
and the US National Science Foundation under grants
PECASE PHY-0643150 and PHY-0548311. JT is supported by the U.S. Department of State through a Fulbright Award.

\newpage
\appendix
\section{The Nine-Point N$^2$MHV Tree Amplitude}\label{9ptappendix}

\begin{table}[h!]\vspace{-0cm}
\mbox{\hspace{-0.7cm}\includegraphics[scale=0.855]{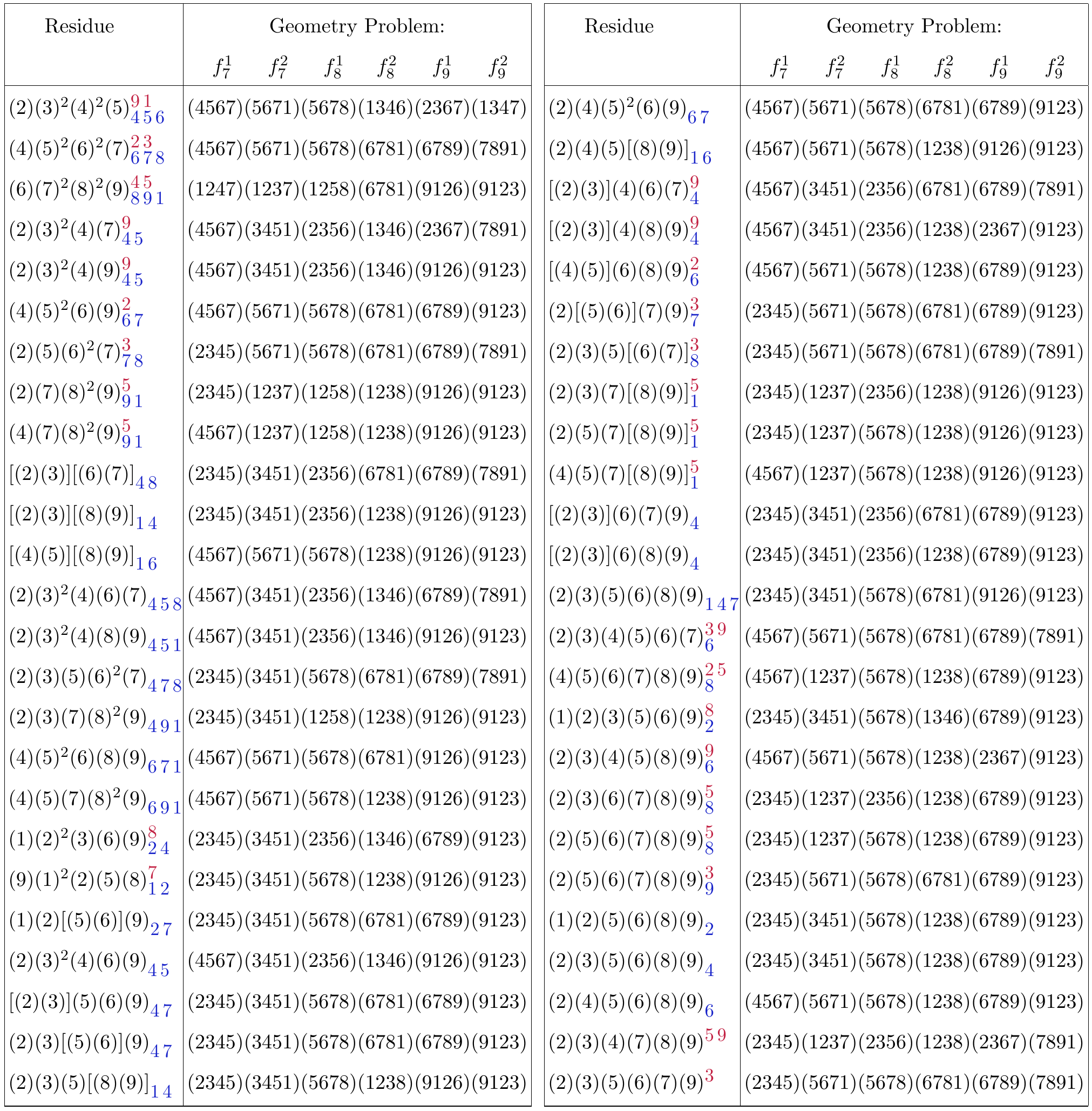}}\end{table}

\newpage
%

\begin{thebibliography}{10}

\bibitem{ArkaniHamed:2009dn}
N.~Arkani-Hamed, F.~Cachazo, C.~Cheung, and J.~Kaplan, ``{A Duality for The
  S-Matrix},''
\href{http://arxiv.org/abs/0907.5418}{{ arXiv:0907.5418 [hep-th]}}.

\bibitem{DG}
L.~Dolan and P.~Goddard, ``{Gluon Tree Amplitudes in Open Twistor String
  Theory},'' \href{http://dx.doi.org/10.1088/1126-6708/2009/12/032}{{\em JHEP}
  {\bf 12} (2009)  032},
\href{http://arxiv.org/abs/0909.0499}{{ arXiv:0909.0499 [hep-th]}}.

\bibitem{NVW}
D.~Nandan, A.~Volovich, and C.~Wen, ``{A Grassmannian \'{E}tude in NMHV
  Minors},''
\href{http://arxiv.org/abs/0912.3705}{{ arXiv:0912.3705 [hep-th]}}.

\bibitem{ABCTtwo}
N.~Arkani-Hamed, J.~Bourjaily, F.~Cachazo, and J.~Trnka, ``{Unification of
  Residues and Grassmannian Dualities},''
\href{http://arxiv.org/abs/0912.4912}{{ arXiv:0912.4912 [hep-th]}}.

\bibitem{SV}
M.~Spradlin and A.~Volovich, ``{From Twistor String Theory To Recursion
  Relations},'' \href{http://dx.doi.org/10.1103/PhysRevD.80.085022}{{\em Phys.
  Rev.} {\bf D80} (2009)  085022},
\href{http://arxiv.org/abs/0909.0229}{{ arXiv:0909.0229 [hep-th]}}.

\bibitem{Mason:2009qx}
L.~Mason and D.~Skinner, ``{Dual Superconformal Invariance, Momentum Twistors
  and Grassmannians},''
  \href{http://dx.doi.org/10.1088/1126-6708/2009/11/045}{{\em JHEP} {\bf 11}
  (2009)  045},
\href{http://arxiv.org/abs/0909.0250}{{ arXiv:0909.0250 [hep-th]}}.

\bibitem{ArkaniHamed:2009vw}
N.~Arkani-Hamed, F.~Cachazo, and C.~Cheung, ``{The Grassmannian Origin of Dual
  Superconformal Invariance},''
\href{http://arxiv.org/abs/0909.0483}{{ arXiv:0909.0483 [hep-th]}}.

\bibitem{Bullimore:2009cb}
M.~Bullimore, L.~Mason, and D.~Skinner, ``{Twistor-Strings, Grassmannians and
  Leading Singularities},''
\href{http://arxiv.org/abs/0912.0539}{{ arXiv:0912.0539 [hep-th]}}.

\bibitem{Kaplan:2009mh}
J.~Kaplan, ``{Unraveling $\mathcal{L}_{n,k}$: Grassmannian Kinematics},''
\href{http://arxiv.org/abs/0912.0957}{{ arXiv:0912.0957 [hep-th]}}.

\bibitem{ABCTone}
N.~Arkani-Hamed, J.~Bourjaily, F.~Cachazo, and J.~Trnka, ``{Local Spacetime
  Physics from the Grassmannian},''
\href{http://arxiv.org/abs/0912.3249}{{ arXiv:0912.3249 [hep-th]}}.

\bibitem{Drummond:2010qh}
J.~M. Drummond and L.~Ferro, ``{Yangians, Grassmannians and T-{D}uality},''
\href{http://arxiv.org/abs/1001.3348}{{ arXiv:1001.3348 [hep-th]}}.

\bibitem{Drummond:2010uq}
J.~M. Drummond and L.~Ferro, ``{The Yangian {O}rigin of the Grassmannian
  {I}ntegral},''
\href{http://arxiv.org/abs/1002.4622}{{ arXiv:1002.4622 [hep-th]}}.

\bibitem{Korchemsky:2010ut}
G.~P. Korchemsky and E.~Sokatchev, ``{Superconformal {I}nvariants for
  {S}cattering {A}mplitudes in $\mathcal{N}=4$ SYM {T}heory},''
\href{http://arxiv.org/abs/1002.4625}{{ arXiv:1002.4625 [hep-th]}}.

\bibitem{DGtwo}
L.~Dolan and P.~Goddard, ``{General Split Helicity Gluon Tree Amplitudes in
  Open Twistor String Theory},''
\href{http://arxiv.org/abs/1002.4852}{{ arXiv:1002.4852 [hep-th]}}.

\bibitem{Broedel:2010rr}
J.~Broedel and S.~He, ``{Dual {C}onformal {C}onstraints and {I}nfrared
  {E}quations from {G}lobal {R}esidue {T}heorems in $\mathcal{N}=4$ SYM
  {T}heory},''
\href{http://arxiv.org/abs/1004.2400}{{ arXiv:1004.2400 [hep-th]}}.

\bibitem{Britto:2004ap}
R.~Britto, F.~Cachazo, and B.~Feng, ``{New Recursion Relations for Tree
  Amplitudes of Gluons},''
  \href{http://dx.doi.org/10.1016/j.nuclphysb.2005.02.030}{{\em Nucl. Phys.}
  {\bf B715} (2005)  499--522},
\href{http://arxiv.org/abs/hep-th/0412308}{{ arXiv:hep-th/0412308}}.

\bibitem{Britto:2005fq}
R.~Britto, F.~Cachazo, B.~Feng, and E.~Witten, ``{Direct Proof of Tree-Level
  Recursion Relation in Yang- Mills Theory},''
  \href{http://dx.doi.org/10.1103/PhysRevLett.94.181602}{{\em Phys. Rev. Lett.}
  {\bf 94} (2005)  181602},
\href{http://arxiv.org/abs/hep-th/0501052}{{ arXiv:hep-th/0501052}}.

\bibitem{Witten:2003nn}
E.~Witten, ``{Perturbative Gauge Theory as a String Theory in Twistor Space},''
  \href{http://dx.doi.org/10.1007/s00220-004-1187-3}{{\em Commun. Math. Phys.}
  {\bf 252} (2004)  189--258},
\href{http://arxiv.org/abs/hep-th/0312171}{{ arXiv:hep-th/0312171}}.

\bibitem{Roiban:2004yf}
R.~Roiban, M.~Spradlin, and A.~Volovich, ``{On the Tree-Level S-Matrix of
  Yang-Mills Theory},''
  \href{http://dx.doi.org/10.1103/PhysRevD.70.026009}{{\em Phys. Rev.} {\bf
  D70} (2004)  026009},
\href{http://arxiv.org/abs/hep-th/0403190}{{ arXiv:hep-th/0403190}}.

\bibitem{RSV}
R.~Roiban, M.~Spradlin, and A.~Volovich, ``{A Googly Amplitude from the B-Model
  in Twistor Space},'' {\em JHEP} {\bf 04} (2004)  012,
\href{http://arxiv.org/abs/hep-th/0402016}{{ arXiv:hep-th/0402016}}.

\bibitem{Roiban:2004ka}
R.~Roiban and A.~Volovich, ``{All Googly Amplitudes from the B-Model in Twistor
  Space},'' \href{http://dx.doi.org/10.1103/PhysRevLett.93.131602}{{\em Phys.
  Rev. Lett.} {\bf 93} (2004)  131602},
\href{http://arxiv.org/abs/hep-th/0402121}{{ arXiv:hep-th/0402121}}.

\bibitem{Griffiths:1978a}
P.~Griffiths and J.~Harris, {\em Principles of Algebraic Geometry}.
\newblock Wiley, New York, 1978.

\bibitem{ABCCKT:2010}
N.~Arkani-Hamed, J.~L. Bourjaily, F.~Cachazo, C.~Cheung, J.~Kaplan, and
  J.~Trnka. Work in progress, 2010.

\end{thebibliography}
%

\providecommand{\href}[2]{#2}\begingroup\raggedright\endgroup

\end{document}